\documentclass[final,preprint,aps,showpacs,fleqn,nofootinbib]{revtex4-2}
\usepackage[toc]{appendix}
\usepackage{epsfig,graphicx}
\graphicspath{./}
\usepackage{amsmath}
\usepackage{hyperref}
\hypersetup{colorlinks,linkcolor=red, citecolor=green}

\newcommand{\tr}{{\rm tr}}

\newcommand{\ave}[1]{\langle #1\rangle}

\newcommand{\re}[1]{(\ref{#1})}
\usepackage[usenames,dvipsnames]{color}

\begin{document}


\title{ILM gluons in perturbative QCD}

\author{M.~Musakhanov, N.~Rakhimov }
\affiliation{Theoretical Physics Department, National University of Uzbekistan, Tashkent 100174, 
Uzbekistan}
\email{musakhanov@gmail.com}

\date\today

\begin{abstract}
In this paper we extend our previous work on gluon propagator in the Instanton Liquid Model (ILM) of the QCD vacuum. This objects presents a lot of interest for studies of the heavy quarkonium $Q\bar Q$ observables in the framework of potential Nonrelativistic QCD (pNRQCD).
Our goal is to evaluate the gluon polarization operator in ILM,  and understand if it gets contributions from infrared (IR) renormalons. We perform a systematic analyis, taking into account both perturbative and nonperturbative effects, and making a double series expansion  in terms
of the strong coupling $\alpha_s(\rho)\sim 0.5$ 
(the scale is given by average instanton size $\rho\approx1/3$~fm)  
and the instanton gas packing fraction $\lambda=\rho^4/R^4\sim 0.01$ 
($R\approx 1$~fm is average inter-instanton distance). 
We demonstrate that there are no IR renormalon related to ILM gluon propagator, 
since instantons generate a ILM gluon dynamical mass.

\end{abstract}

\maketitle

\thispagestyle{empty}

\section{Introduction}

\begin{figure}[hbt]
\begin{centering}
\includegraphics[scale=0.8]{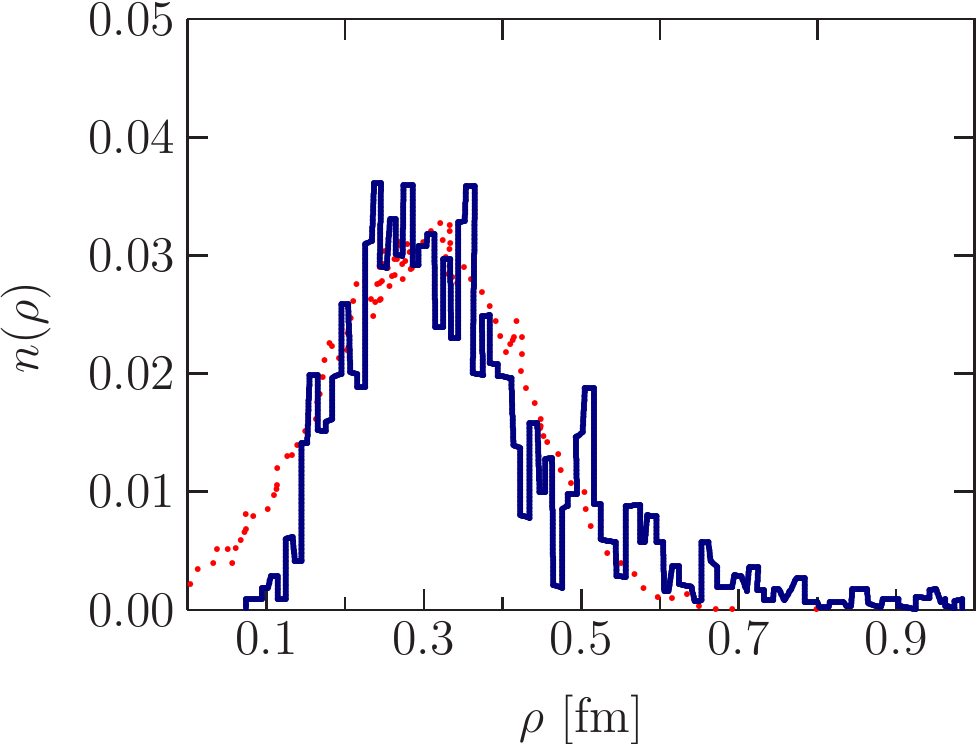} 
\par\end{centering}
\caption{The dependence of the instanton size distribution function $n$  on
the instanton size parameter $\rho$. The dots correspond to the calculations
in the framework of ILM while the continuous lines correspond to the
lattice simulations~\cite{Millo:2011zn}.}
\label{instantonsize} 
\end{figure}

The quasi-classical approach to QCD  establishes importance of the
topologically non-trivial classical solutions of chromodynamics in
Euclidean space \textendash{} instantons~\cite{Belavin:1975fg,Pol,tH,tH1}.
They have a quantum meaning of the paths in the internal Chern-Simons
space, connecting classical vacuum states with different Chern-Simons
numbers~\cite{FJR1976,Jackiw:1976pf}. Accordingly in quantum mechanics
these paths correspond to tunneling processes between different classical
vacuum states of chromodynamics. On the base of these ideas, it was formulated the
Instanton Liquid Model (ILM) for the QCD vacuum 
(see the reviews~\cite{Schafer:1996wv,shuryak2018,Diakonov:2002fq}).
In ILM framework the four-dimensional Euclidean space-time is populated
by randomly distributed QCD instantons and anti-instantons. Their
sizes and densities are controlled by instanton-instanton and instanton-antiinstanton
interactions.  The average instanton size $\bar{\rho}$ and average
inter-instanton distance $\bar{R}$ have been independently estimated
using different variational, phenomenological and numerical
methods, yielding $\bar{\rho}\approx1/3$~fm, and $\bar{R}\approx$~fm.
These values were confirmed by lattice measurements~\cite{Chu:1994vi,Negele:1998ev,DeGrand:2001tm,Faccioli:2003qz}.
The instanton size distribution $n(\rho)$ has been studied by the
lattice simulations~\cite{Millo:2011zn} (see Fig.\ref{instantonsize}).

The main success of ILM framework in the last years was a clear explanation
of the Spontaneous Breaking of Chiral Symmetry~\cite{Diakonov:1987ty,Diakonov:1985eg}, 
including Chiral Perturbation
Theory results of light quarks physics~\cite{Goeke:2007bj,Kim:2004hd,Kim:2005jc}. Further extension
of ILM \textendash{} Dyon Liquid Model framework even provided a possible way to understanding
of the confinement ~\cite{Kraan:1998kp,Kraan:1998pm,Lee:1998bb,Diakonov:2009jq,Liu:2015ufa,Liu:2015jsa}.


\textbf{Details of Instanton Liquid Model for QCD Vacuum.}

In this section we will remind briefly the details of ILM approach
(see the reviews~\cite{Diakonov:2002fq,Schafer:1996wv,shuryak2018}
and references therein for more details). These results will be used
below for the calculations.

Due to diluteness of the instanton gas,  the background field in ILM
approach  is given by a simple sum of instanton and antiinstanton
fields, 
\begin{equation}
A(\xi)=\sum_{i}A_{i}(\xi_{i})
\end{equation}
 where $\xi_{i}=(z_{i},U_{i},\rho_{i})$ are collective coordinates
of instantons. It is also necessary to note that the instanton field
has a specific strong coupling dependence given by $A\sim1/g$. The
quaitization of the gluonic field in the instanton background is done
extending the total field is $A(\xi)+a,$ where the quantum fluctuation
$a$ might be treated perturbatively. In what follows we will need
to average over all collective degrees of freedom of instantons  $\xi=(\xi_{1},.....,\xi_{N})$,
and we will use for this a shorthand notation 
\begin{eqnarray}
\ave{...}_{\xi}=\int D\xi...,\,\,\,\,\int D\xi=1.
\end{eqnarray}
The perturbative evaluation of the loop corrections introduces dependence
on the regularization scale $\mu$, which determines the magnitude
of the gauge coupling constant given at that scale $\alpha_{s}(\mu)$.
This dependence might be related to the scale of strong interactions
$\Lambda$  given by 
\begin{eqnarray}
\Lambda & = & \mu\,\exp\left(-\frac{2\pi}{b_{1}\alpha_{s}(\mu)}\right)\,\left(\frac{4\pi}{b_{1}\alpha_{s}(\mu)}\right)^{b_{2}/2b_{1}^{2}}\,\left(1+O\left(\alpha_{s}\right)\right),\label{Lambda}\\
b_{1} & = & \frac{11}{3}N_{c}-\frac{2}{3}N_{f},\qquad b_{2}=\frac{34}{3}N_{c}^{2}-\frac{13}{3}N_{c}N_{f}+\frac{N_{f}}{N_{c}},\label{b}
\end{eqnarray}
where $N_{c}=3$ is the number of quark colors and $N_{f}$ is the
number of acting quark flavors, while $b_{1},\,\,b_{2}$ are the coefficients
of QCD $\beta$-function defined as 
\begin{eqnarray}
\mu^{2}\frac{d\alpha_{s}(\mu)}{d\mu^{2}}=\sum_{k=0}^{\infty}\beta_{k}\alpha_{s}^{k+2}=-\frac{1}{4\pi}b_{1}\alpha_{s}^{2}(\mu)-\frac{1}{(4\pi)^{2}}b_{2}\alpha_{s}^{3}(\mu)+...,\,\beta_{0}=-\frac{1}{4\pi}b_{1},\,\beta_{1}=-\frac{1}{4\pi}b_{2};\label{beta-function}
\end{eqnarray}
For many practical applications the relevant values of normalization
scale $\mu$ significantly exceed the soft scale $\Lambda$.  For this
reason, using smallness of $\alpha_{s}(\mu)$, it is possible to rewrite
$\alpha_{s}(\mu)$ in a conventional asymptotic form 
\begin{eqnarray}
\frac{2\pi}{\alpha_{s}(\mu)}=b_{1}\ln\frac{\mu}{\Lambda}+\frac{b_{2}}{2b_{1}}\ln\ln\frac{\mu^{2}}{\Lambda^{2}}+O\left(\frac{1}{\ln\frac{\mu}{\Lambda}}\right).\label{alpha_s}
\end{eqnarray}
 It is natural to expect that all dimensional physical observables
in QCD are proportional to $\Lambda$ in the appropriate power. In
the ILM approach the natural regulaization scale is set by the instanton
size $\rho$, so the coupling $\alpha_{s}(\rho)$, which controls
the dynamics of strong interactions in the instanton background, is
given to the 1-loop accuracy by 
\begin{eqnarray}
2\pi/\alpha_{s}^{(1)}(\rho)=b_{1}\ln\frac{1}{\Lambda\rho},
\end{eqnarray}
whereas inclusion of 2-loop corrections modifies it as 
\begin{eqnarray}
2\pi/\alpha_{s}^{(2)}(\rho)=b_{1}\ln\frac{1}{\Lambda\rho}+\frac{b_{2}}{2b_{1}}\ln{\ln\frac{1}{\Lambda^{2}\rho^{2}}}\label{alphas2}
\end{eqnarray}
We need to mention that the estimates for the scale $\Lambda$ depend
on the accepted regularization scheme: for example, in $\overline{{\rm MS}}$
scheme its values are slightly smaller tham in Pauli-Villars regularization,
$\Lambda_{\overline{{\rm MS}}}=e^{-\frac{1}{22}}\Lambda=0.955\Lambda.$
The distribution over the sizes of instantons (``instanton weight
function'') in two-loop approximation is given by

\begin{eqnarray}
d_{0}(\rho) & = & \frac{C(N_{c})}{\rho^{5}}\left(\frac{2\pi}{\alpha_{s}^{(1)}(\rho)}\right)^{2N_{c}}\exp\left[-\frac{2\pi}{\alpha_{s}^{(2)}(\rho)}+\left(2N_{c}-\frac{b_{2}}{2b_{1}}\right)\!\frac{b_{2}}{2b_{1}}\frac{\alpha_{s}^{(1)}(\rho)\ln(2\pi/\alpha_{s}^{(1)}(\rho))}{2\pi}\right]\nonumber \\
 & \sim & \frac{1}{\rho^{5}}(\Lambda\rho)^{(\frac{11}{3}N_{c}-\frac{2}{3}N_{f})},\label{d02}
\end{eqnarray}
and clearly is divergent ar large $\rho.$ This diveregence is a mere
consequence of dilute gas approximation and disappears when the inter-instantons
interactions are taken into account. We need to mention that for large
instantons the strong coupling increases drasticaly, for this reason
it is complicated to evaluate this modification from the first principles.
However, the estimates based  on variational principle suggest that
suppression of large-size dipoles is quite fast and might be described
by a Gaussian cutoff, 
\begin{eqnarray}
d_{0}(\rho)\to d(\rho)=d_{0}(\rho)\,\exp\left(-c\,\rho^{2}/R^{2}\right),\label{gauss}
\end{eqnarray}
where $c$ is some constant. In fact, the function $d(\rho)$  is
a rather narrow distribution peaked around $\bar{\rho}$ (\ref{rhoav});
therefore for practical estimates we may just neglect the width of
this distribution. In what follows we will use for our 
estimates the average instanton size and the average separation between
instantons~\cite{Diakonov:2002fq}
\begin{eqnarray}
\bar{\rho}\simeq0.48/\Lambda_{\overline{{\rm MS}}}\simeq0.35\,{\rm fm},\,\,\,\,\,\bar{R}\simeq1.35/\Lambda_{\overline{{\rm MS}}}\simeq0.95\,{\rm fm},\label{rhoav}
\end{eqnarray}
where the scale $\Lambda_{\overline{{\rm MS}}}=280\,{\rm MeV}$ is
extracted from phenomenological studies of strong coupling. These
values agree with estimates from the lattice~\cite{Chu:1994vi,Negele:1998ev,DeGrand:2001tm,Faccioli:2003qz},
as well as  phenomenological applications of instantons~\cite{Schafer:1996wv}.

\textbf{Application of ILM to light quark physics.}

The ILM framework provides a very natural nonperturbative explanation
of the  Spontaneous Breaking of the Chiral Symmetry (SBCS) in QCD
(see ~\cite{Schafer:1996wv,Diakonov:2002fq} for the review), and
as a consequence provides a consistent framework for microscopic description
of the pions, giving the possibility to evaluate microscopically all
the couplings in chiral lagrangians. Technically, the ILM approach
the possibility to explain SBCS and the goldstone nature of the pion
are closely related to the fact that the dynamics of light quarks
in the instanton background is strongly affected by the  presence
of zero-modes  in light quark propagator.  A consistent way for the
derivation of light quark determinant and on this base the light quark
partition function was proposed in the number of works~\cite{Goeke:2007bj,Kim:2004hd,Kim:2005jc,Musakhanov:1998wp,Musakhanov:2002vu,Musakhanov:2002xa}. 
In what follows we will extend this approach for analysis of hadrons
involving heavy quarks.

\textbf{Radiative corrections to gluon propagator in QCD.}

 While formally in the heavy quark mass limit we could expect that
the quarkonia might be described perturbatively, the for practical
applications, especially in charm sector, the numerical values of
$\alpha_{s}$ still are quite significant. For this reason a successful
application of potential Non-Relativistic QCD (pNRQCD) to heavy
quarks physics requires to take into account higher order correcions
in  $\alpha_{s}$ , as could be evidenced from analysis of the so
called IR renormalons problem (see recent work~\cite{Beneke:2021lkq}
and the references therein). While at short distances the interaction
potential between quarks is still dominated by the one-gluon exchange,
we understand that behaviour of the running strong coupling $\alpha_{s}(q)$
at small momentum becomes more pronounced.

The present study is the extension of our previous work on gluon propagator
in ILM~\cite{hutter,Musakhanov:2017erp}, which allowed to evaluate
the dynamical momentum-dependent gluon mass  $M_{g}(q)$. Furthermore,
ILM gluon propagator was applied for the calculations of lowest order
on strong coupling $\alpha_{s}$ one-gluon exchange potential in addition
to direct instanton contribution potential for the problem of heavy
quarkonium in QCD~\cite{Musakhanov:2020hvk}. Our aim is to calculate
gluon polarization operator in ILM and understand if there are IR
renormalons in the $Q\bar{Q}$ observables. A systematic analysis
including both perturbative and nonperturbative effects requires a
double expansion series in terms of $\alpha_{s}(\rho)\sim0.5$ and
$\lambda\sim0.01$. In order to perform such an analysis we assume
that $\alpha_{s}\sim\lambda^{1/4}$ which is quite reasonable according
to the phenomenological studies.

\section{Radiative corrections to gluon propagator in ILM}

In QED the lowest order polarization operator certainly is related
only to fermion one-loop Feynman diagrams, while in QCD we have also
gluon and ghost contributions. In QCD insertion of quark loops leads
to the $N_{f}$ depended part in $b_{1}$ Eq.~\re{b}. The substitution
of the QCD full $b_{1}$ in quark loops-chain diagrams is referred
to as ``non-abelianization''~\cite{Broadhurst:1994se,Beneke:1994qe}. 

In ILM it is natural to split the light quark determinant into the
low- and high-frequency parts according to ${\rm Det}={\rm Det}_{{\rm high}}\times{\rm Det}_{{\rm low}}$
(see the reviews~\cite{Schafer:1996wv,Diakonov:2002fq}) and concentrate
on the evaluation of ${\rm Det}_{{\rm low}}$ , which is responsible
for the low-energy domain. The high-energy part ${\rm Det}_{{\rm high}}$
is responsible mainly for the perturbative coupling renormalization
discussed above. As was demonstrated before in our previous papers~\cite{Musakhanov:1998wp,Goeke:2007bj},
a proper inclusion of current quark mass and external fields needs
some care and leads to the fermionic representation of ${\rm Det}_{{\rm low}}$
in the presence some external vector field $a_{\mu}$ as 
\begin{eqnarray}
{\rm Det}_{{\rm low}}=\int\prod_{f}D\psi_{f}D\psi_{f}^{\dagger}\exp\left(\int\sum_{f}\psi_{f}^{\dagger}(\hat{p}\,+\,\hat{a}\,+\,im)_{f}\psi_{f}\right)\prod_{\pm}^{N_{\pm}}\tilde{V}_{\pm,f}[\xi,\psi^{\dagger},\psi,a]\;,\label{det}
\end{eqnarray}
where 
\begin{eqnarray}
\tilde{V}_{\pm}[\xi,\psi^{\dagger},\psi,a]=\int d^{4}x\left(\psi^{\dagger}(x)\,\bar{L}^{-1}(x,z_{\pm})\,\hat{p}\Phi_{\pm,0}(x;\xi_{\pm})\right)\int d^{4}y\left(\Phi_{\pm,0}^{\dagger}(y;\xi_{\pm})(\hat{p}\,L^{-1}(y,z_{\pm})\psi(y)\right),\label{tildeV}
\end{eqnarray}
$\Phi_{\pm,0}$ are the light quarks zero-modes and the gauge links
$L_{i}$ are defined as 
\begin{eqnarray}
L_{i}(x,z_{i})={\rm P}\exp\left(i\int_{z_{i}}^{x}dy_{\mu}a_{\mu}(y)\right),\,\,\,\bar{L}_{i}(x,z_{i})=\gamma_{4}L_{i}^{\dagger}(x,z_{i})\gamma_{4}.\label{transporter}
\end{eqnarray}
The partition function in ILM $Z[j]$ (normalized as $Z[0]=1$)  is
given by 
\begin{align}
Z[j] & =\frac{1}{\ave{{\rm Det}_{{\rm low}}[\xi,m]}_{\xi}}\int D\xi\,Da\,{\rm Det}_{{\rm low}}[\xi,a,m]e^{-[S_{eff}[a,A(\xi)]+(ja)]}\nonumber \\
 & =\frac{1}{\ave{{\rm Det}_{{\rm low}}[\xi,m]}_{\xi}}\int D\xi\,Da\,{\rm Det}_{{\rm low}}[\xi,\frac{\delta}{\delta j_{\mu}},m]e^{-[S_{eff}[a,A(\xi)]+(ja)]}\nonumber \\
 & \approx\frac{1}{\ave{{\rm Det}_{{\rm low}}[\xi,m]}_{\xi}}\int D\xi\,{\rm Det}_{{\rm low}}[\xi,\frac{\delta}{\delta j_{\mu}},m]e^{\frac{1}{2}(j_{\mu}S_{\mu\nu}(\xi)j_{\nu})},\label{Z}
\end{align}
where $a_{\mu}$are perturbative gluons (quantum fluctuations around
instanton background) introduced earlier, $j_{\mu}$ are their external
sources, and light quarks contribute  via their determinant~\re{det}.
We also  used the shorthan notation $(ja)=\int d^{4}xj_{\mu}^{a}(x)a_{\mu}^{a}(x).$
The measure of integration in ILM is given explicitly as $D\xi=\prod_{i}d\xi_{i}=V^{-1}\prod_{i}dz_{i}dU_{i}$,
and the integration over the  instantons' sizes $\rho_{i}$ is disregarded
in view of the above-mentioned smallness of the width of the instanton
distribution.

In order to simplify further discussion, temporarily we will replace
the real gluon field $a_\mu$ with a scalar \char`\"{}gluon\char`\"{}
field $\phi$. This allows us to suppress the  gauge links $L_{i}$.
Furthermore, in Eq.~\re{det} we will change $\phi$ to $\frac{\delta}{\delta j}$,
so the partition function might be rewritten as
\begin{eqnarray}
Z[j]=\frac{1}{\ave{{\rm Det}_{{\rm low}}[\xi,m]}_{\xi}}\int D\xi\,D\phi\,{\rm Det}_{{\rm low}}[\xi,\frac{\delta}{\delta j},m]e^{-[S_{eff}[\phi,\xi]+(j\phi)]}\label{Z1}\\
=\frac{1}{\ave{{\rm Det}_{{\rm low}}[\xi,m]}_{\xi}}\int D\xi\,{\rm Det}_{{\rm low}}[\xi,\frac{\delta}{\delta j},m]e^{\frac{1}{2}(j\Delta j)}\nonumber 
\end{eqnarray}
The scalar \char`\"{}gluon\char`\"{} propagator $\Delta$ in background
field $A$  of the instanton gas is given by 
\begin{eqnarray}
 &  & \Delta=(p+A)^{-2}=(p^{2}+\sum_{i}(\{p,A_{i}\}+A_{i}^{2})+\sum_{i\neq j}A_{i}A_{j})^{-1},\,\,\,\Delta_{0}=p^{-2},\\
 &  & \tilde{\Delta}=(p^{2}+\sum_{i}(\{p,A_{i}\}+A_{i}^{2}))^{-1},\,\,\,\,\Delta_{i}=P_{i}^{-2}=(p^{2}+\{p,A_{i}\}+A_{i}^{2})^{-1}.\nonumber 
\end{eqnarray}
where $\Delta_{0}$ is the free propagator, $\Delta_{i}$
is the propagator in the field of a single instanton $i$, and $\tilde{\Delta}$
is the propagator in the field of instanton gas in dilute approximation
(when overlap of the neighbour instantons is neglected)
. There {are}
no zero modes in $\Delta_{i}^{-1}=P_{i}^{2}$ and $\Delta^{-1}=P^{2}$,
{which} means {the} existence of the inverse operators $\Delta_{i}$
and $\Delta$. Now we would like to discuss evaluation of ${\tilde{\Delta}}.$
Expanding it over $(\{p,A_{i}\}+A_{i}^{2})$ {carrying out further
re-summation}, {we obtain} the multi-scattering series 
\begin{eqnarray}
\tilde{\Delta}=\Delta_{0}+\sum_{\pm}(\Delta_{\pm}(\xi_{\pm})-\Delta_{0})+...,
\end{eqnarray}
where the expansion is done over the packing fraction $\lambda=\rho^{4}/R^{4}\sim0.01.$,
which in essence characterizes the fraction of 4D space occupied by
instantons. The difference between exact and dilute gas approximation
propagators is suppressed in this limit, $\Delta=\tilde{\Delta}+O(\lambda^{2}),$
so, the partition function \re{Z1} might be rewritten as 
\begin{eqnarray}
Z[j]=\frac{1}{Z[0]}\int\prod_{f}D\psi_{f}D\psi_{f}^{\dagger}\exp\left(\sum_{f}\psi_{f}^{\dagger}(\hat{p}\,+\,g\frac{\delta}{\delta j}\,+\,im)\psi_{f}\right)\exp{[\frac{1}{2}(j\Delta_{0}j)]}\\
\times\prod_{\pm}^{N_{\pm}}\left(\int d\xi_{\pm}\exp{[\frac{1}{2}(j(\Delta_{\pm}(\xi_{\pm})-\Delta_{0})j)]}\prod_{f}V_{\pm,f}[\xi_{\pm},\psi^{\dagger},\psi]\right)\nonumber \\
=\frac{1}{Z[0]}\int\prod_{f}D\psi_{f}D\psi_{f}^{\dagger}\exp\left(\sum_{f}\psi_{f}^{\dagger}(\hat{p}\,+\,g\frac{\delta}{\delta j}\,+\,im)\psi_{f}\right)\exp{[\frac{1}{2}(j\Delta_{0}j)]}\nonumber \\
\times\prod_{\pm}\left(\int d\xi_{\pm}\exp{[\frac{1}{2}(j(\Delta_{\pm}(\xi_{\pm})-\Delta_{0})j)]}\prod_{f}V_{\pm,f}[\xi_{\pm},\psi^{\dagger},\psi]\right)^{N_{\pm}}\nonumber 
\end{eqnarray}
We may rewrite the last term in the bracket as 
\begin{eqnarray}
 &  & \ave{\exp{[\frac{1}{2}(j(\Delta_{\pm}(\xi_{\pm})-\Delta_{0})j)]}\prod_{f}V_{\pm,f}[\xi,\psi^{\dagger},\psi]}_{\xi}\\
 &  & =\ave{\exp{[\frac{1}{2}(j(\Delta_{\pm}(\xi_{\pm})-\Delta_{0})j)]}}_{\xi}\ave{\prod_{f}V_{\pm,f}[\xi,\psi^{\dagger},\psi]}_{\xi}\nonumber \\
 &  & +\left(\ave{\exp{[\frac{1}{2}(j(\Delta_{\pm}(\xi_{\pm})-\Delta_{0})j)]}\prod_{f}V_{\pm,f}[\xi,\psi^{\dagger},\psi]}_{\xi}\right.\nonumber \\
 &  & \left.-\ave{\exp{[\frac{1}{2}(j(\Delta_{\pm}(\xi_{\pm})-\Delta_{0})j)]}}_{\xi}\ave{\prod_{f}V_{\pm,f}[\xi,\psi^{\dagger},\psi]}_{\xi}\right)\nonumber 
\end{eqnarray}
We see that the integration over $\xi_{\pm}$ in the second term leads
to the interaction terms between \char`\"{}gluons\char`\"{} and light
quarks. For a moment we will neglect this contribution. Furthermore,
we will neglect the \char`\"{}gluon\char`\"{}-\char`\"{}gluon\char`\"{}
interactions generated by instantons, which appear due to integration
over $\xi_{\pm}$,  
\begin{eqnarray}
\ave{\exp{[\frac{1}{2}(j(\Delta_{\pm}(\xi_{\pm})-\Delta_{0})j)]}}_{\xi}\approx\exp{[\frac{1}{2}(j\ave{(\Delta_{\pm}(\xi_{\pm})-\Delta_{0})}_{\xi}\,j)]}
\end{eqnarray}

Now we may exponentiate $V^{N}$ by using Stirling-like formula 
\begin{eqnarray}
V^{N}=\int d\eta\exp(N\ln\frac{N}{\eta}-N+\eta V),
\end{eqnarray}
in order to rewrite the partition function $Z$ as
\begin{eqnarray}\label{ZNf}
Z[j]\approx\frac{1}{Z[0]}\int\prod_{\pm}d\eta_{\pm}\prod_{f}D\psi_{f}D\psi_{f}^{\dagger}\exp\left(\sum_{f}\psi_{f}^{\dagger}(\hat{p}\,+\,g\frac{\delta}{\delta j}\,+\,im)\psi_{f}\right)\\
\times\exp\sum_{\pm}\left(N_{\pm}\ln\frac{N_{\pm}}{\eta_{\pm}}-N_{\pm}+\eta_{\pm}\ave{\prod_{f}V_{\pm,f}[\xi,\psi^{\dagger},\psi]}_{\xi}\right)\nonumber \\
\times\exp{[\frac{1}{2}(j\Delta_{0}j)]}\,\exp{[\frac{1}{2}j\sum_{\pm}N_{\pm}\ave{(\Delta_{\pm}(\xi_{\pm})-\Delta_{0})}_{\xi}\,j]},\nonumber 
\end{eqnarray}
where in the last string we see ILM \char`\"{}gluon\char`\"{} propagator
\[
\bar\Delta=\Delta_{0}+\sum_{\pm}N_{\pm}\ave{(\Delta_{\pm}(\xi_{\pm})-\Delta_{0})}_{\xi}+O(\lambda^{2}).
\]
For a moment we'll consider  a theory with just a single quark flavour
$N_{f}=1$ and equal number of instantons and antiinstantons $N_{\pm}=N/2$.
The integration over $\eta_{\pm}$ at saddle-point approximation yields $\eta_{\pm}=\eta$,
so we may get
\begin{eqnarray}\label{Nf=1}
 &  & \int\prod_{\pm}d\eta_{\pm}D\psi D\psi^{\dagger}\exp\left(\psi^{\dagger}(\hat{p}\,+\,g\frac{\delta}{\delta j}\,+\,im)\psi\right)\\
 &  & \times\exp\sum_{\pm}\left(N_{\pm}\ln\frac{N_{\pm}}{\eta_{\pm}}-N_{\pm}+\eta_{\pm}\ave{V_{\pm}[\xi,\psi^{\dagger},\psi]}_{\xi}\right)\nonumber \\
 &  & =\exp\left[{\rm Tr}\ln\left(\hat{p}\,+\,g\frac{\delta}{\delta j}\,+i(m+M(p))\right)+N\ln\frac{N/2}{\lambda}-N\right],\nonumber \\
 &  & N={\rm Tr}\frac{iM(p)}{\hat{p}\,+\,i(m+M(p))},\,\,\,M(p)=\frac{\eta}{N_{c}}(2\pi\rho F(p))^{2},\,\,\,F(q)=q\rho K_{1}(q\rho).\label{SP1}
\end{eqnarray}
where $M(p)$ is the dynamical (constituent) quark mass. The partition
function in this approximation becomes: 
\begin{eqnarray}
Z[j]\approx\frac{1}{Z[0]}\exp\left[{\rm Tr}\ln\left(\hat{p}\,+\,g\frac{\delta}{\delta j}\,+i(m+M(p))\right)+N\ln\frac{N/2}{\lambda}-N\right]\exp{[\frac{1}{2}j\,\bar{\Delta}\,j]}.\label{Z-Nf=1}
\end{eqnarray}
Since 
\begin{eqnarray}
Z[0]=\exp\left[{\rm Tr}\ln\left(\hat{p}\,+\,i(m+M(p))\right)+N\ln\frac{N/2}{\lambda}-N\right]
\end{eqnarray}
the Eq.~(\ref{Z-Nf=1}) might be rewritten as  
\begin{eqnarray}
Z[j]\approx\exp\left[{\rm Tr}\ln\left(1+\,g\frac{\delta}{\delta j}\,(\hat{p}\,+i(m+M(p))^{-1}\right)\right]\exp{[\frac{1}{2}j\,\bar{\Delta}\,j]}.\label{Z-Nf=1-1}
\end{eqnarray}
Next we will consider the case of two quark flavours ($N_{f}=2$).
As earlier, we assume the equality of number of instantons and antiinstantons
($N_{\pm}=N/2$), and integrate over $\eta_{\pm}$ at saddle-point
approximation. In this approximation we may find that  $\eta_{\pm}=\eta$.
For $N_{f}=2$ case the effective action includes a nonlocal 4-quark
interaction vertex. The latter might be rewritten in a simpler form,
making a bosonisation, which essentially replaces the 4-quark interaction
with a new interaction vertices of quarks with scalar and pseudoscalar
fields of different isospin ($\sigma,\,\eta,\,\vec{\sigma},\,\vec{\phi}$).
Due to spontaneous violation of chiral symmetry,  the scalar meson
field $\sigma$ has non-zero vacuum expectation $\sigma_{0}$, and
in what follows we will use notation $\Phi'=(\sigma',\vec{\phi}',\eta',\vec{\sigma}')$
for the quantum fluctuations of this bosonic field around the vacuum
$\sigma_{0}$. Straightforward evaluation shows that similar to $N_{f}=1$
case,  the quarks acquire dynamical (constituent) mass, and its $p$-dependence
is given by an expression similar to~(\ref{SP1}),
\begin{eqnarray}
M(p)=\frac{\eta^{0.5}}{2c}(2\pi\rho)^{2}F^{2}(p)\sigma_{0},\,\,\,c^{2}=\frac{(N_{c}^{2}-1)2N_{c}}{2N_{c}-1}.\,\,
\end{eqnarray}
The magnitude of the mass is controlled by the  non-zero vacuum expectation
$\sigma_{0}$, which might be fixed from the so-called gap equation
\begin{eqnarray}
N=0.5{\rm Tr}\frac{iM(p)}{\hat{p}+im+iM(p)},\,\,\,V\sigma_{0}^{2}={\rm Tr}\frac{iM(p)}{\hat{p}+im+iM(p)}
\end{eqnarray}
where ${\rm Tr}(...)=\tr_{D}\tr_{c}\tr_{f}\int d^{4}x<x|(...)|x>$.
The partition function in this case might be rewritten as an effective
interaction of quarks with mesonic fileds
 $\Phi'=(\sigma',\vec{\phi}',\eta',\vec{\sigma}')$
\begin{eqnarray}
Z[j] & \approx & \frac{1}{Z[0]}\int D\Phi'\prod_{f}D\psi_{f}D\psi_{f}^{\dagger}\exp\left[N/2\ln\frac{N}{2\eta}-N/2-\frac{1}{2}V\sigma_{0}^{2}-\frac{1}{2}\int dx\,{\Phi'}^{2}\right.\label{Z-Nf=2}\\
 &  & \left.+\sum_{f}\psi_{f}^{\dagger}\left(\hat{p}\,+\,g\frac{\delta}{\delta j}\,+\,im_{f}+iM_{f}(p)+\frac{iM}{\sigma_{0}}F(p)\Phi'F(p)\right)\psi_{f}\right]\exp{\left[\frac{1}{2}j\,\bar{\Delta}\,j\right]}\nonumber 
\end{eqnarray}
where ${\Phi'}^{2}={\sigma'}^{2}+{\vec{\phi'}}^{2}+{\vec{\sigma'}}^{2}+{\eta'}^{2}.$ We
may assume that if we neglect by meson fluctuations, at any $N_{f}$
 we may  approximate the action of light quarks by Eq.~\re{Z-Nf=2}.

Up to now we considered the case of scalar \char`\"{}gluons\char`\"{}. The extension
of these results for the case of  real gluons is straightforward and
as was shown in our previous paper~\cite{Musakhanov:2017erp}, yields
for the partition function  
\begin{eqnarray}
Z[j]\approx\exp\left[{\rm Tr}\ln\left(1+\,g\frac{\delta}{\delta j_{\rho}}\,(\hat{p}\,+i(m+M(p))^{-1}\right)\right]\exp{[\frac{1}{2}j_{\mu}\,\bar{S}_{\mu\nu}\,j_{\nu}]},\label{Zreal}
\end{eqnarray}
where we neglected  the gauge links $L_{i}$ contributions (see Eq.~\re{transporter}),
and the ILM gluon propagator is given by
\begin{eqnarray}
\bar{S}_{\mu\nu}(q)=\left(\delta_{\mu\nu}-(1-\xi)\frac{q_{\mu}q_{\nu}}{q^{2}}\right)\frac{1}{q^{2}+M_{g}^{2}(q)},\,\,\,\,M_{g}(q)=F_{g}(q)M_{g},\label{eq:GluonPropagator}\\
M_{g}=[\frac{6\rho^{2}}{(N_{c}^{2}-1)R^{4}}4\pi^{2}]^{1/2},\,\,\,F_{g}(q)=q\rho K_{1}(q\rho).\nonumber 
\end{eqnarray}

It is obvious that Eq.~\re{Z} generate light quarks loops contributions
to the gluon propagator, which can be summed-up to geometrical progression
as: 
\begin{eqnarray}
\bar{S}_{\mu\nu}(q)\frac{1}{1-\frac{q^{2}}{q^{2}+M_{g}^{2}(q)}\pi(q)}\label{barS-with-loops}
\end{eqnarray}
where we used Landau gauge $\xi=0$.The gluon polarization operator
$\pi_{{\mu\nu}}^{ab}(q)=\delta_{ab}(q^{2}\delta_{\mu\nu}-q_{\mu}q_{\nu})\,\pi(q)$
in the lowest order in $\alpha_{s}$ is given by contribution of light
quarks loops, 
\begin{eqnarray}
\pi_{0,\mu\nu}^{ab}(q)=4\pi\alpha_{s}\int\frac{d^{d}p}{(2\pi)^{d}}\tr\,t_{a}\gamma_{\mu}\frac{\hat{p}-i(m+M(p))}{p^{2}+(m+M(p))^{2}}t_{b}\gamma_{\nu}\frac{\hat{p}-\hat{q}-i(m+M(p-q))}{(p-q)^{2}+(m+M(p-q))^{2}};\label{loop}
\end{eqnarray}
where$d=4-\epsilon$ is the dimension in $\overline{{\rm MS}}$ scheme,
$\mu$ is thenormalization point,and we regularized the polarization
operator as $\pi(q)\to\pi(q)-\pi(\mu)$ in order to remove the ultraviolet
logarithmic divergence $(1/\epsilon)$ in Eq.~\re{loop}. Straightforward
evaluation  leads to the standard answer~\cite{Beneke:2021lkq}:
\begin{eqnarray}
\pi_{0}(q)=\beta_{0}\,\alpha_{s}(\mu)\ln\frac{q^{2}e^{-C}}{\mu^{2}}
\end{eqnarray}
where $C=5/3$ in $\overline{{\rm MS}}$ scheme of regularization,
and we use full $\beta_{0}$ which is meaning ``non-abelianization''~\cite{Broadhurst:1994se,Beneke:1994qe}.
Similarly we can consider  radiative correction $\Delta m$ to the
quark mass $m$. For its evaluations we have to use ILM gluon propagator
with radiative corrections Eq.~\re{barS-with-loops}. The calculation
is similar to ~\cite{Beneke:2021lkq} and leads to
\begin{eqnarray}
 &  & \Delta m=4\pi\alpha_{s}(\mu)C_{F}\mu^{2\epsilon}\int\frac{d^{d}q}{(2\pi)^{d}}\frac{\gamma_{\mu}(\hat{p}-\hat{q}-im)\gamma_{\nu}}{(p-q)^{2}+m^{2}}\label{Deltam}\\
 &  & \times\frac{1}{q^{2}+M_{g}^{2}(q)}(\delta_{\mu\nu}-q_{\mu}q_{\nu}/q^{2})\sum_{n=0}^{\infty}\left[\beta_{0}\alpha_{s}(\mu)\frac{q^{2}}{q^{2}+M_{g}^{2}(q)}\ln\left(\frac{q^{2}e^{-C}}{\mu^{2}}\right)\right]^{n}+{\rm counterterms},\nonumber 
\end{eqnarray}
where the color factor $C_{F}=4/3.$ 

{\bf Infrared region contribution to $\Delta m.$}

In the infrared region $(q\leq\mu)$
the dynamical gluon mass $M_{g}(q)$ might be approximated as a constant,
$M_{g}(q)\approx M_{g}(0)\equiv M_{g}$. 

The evaluation of the the
typical integrals in a series~(\ref{Deltam}) in this region yields
 \begin{eqnarray}
\Delta m_{IR}
 =  -\frac{4\pi C_{F}}{\beta_0}(-\beta_0\alpha_s(\mu))
 \sum_{n=0}^\infty (-\beta_0\alpha_s(\mu))^n c_n ,
\label{deltamIR}\\
c_n  = (-1)^n\int_0^\mu dq\, \left(\frac{q^2}{q^2+M_g^2}\right)^{n+1}\ln^n\left(\frac{q^2}{\mu^2}\right)
\\\nonumber
= \mu\,(-1)^n\int_0^1 dx\, \left(\frac{x^2}{x^2+a_g^2}\right)^{n+1}
\ln^{n} x^2
\label{typical}
\end{eqnarray}
where $a_g=M_g/\mu < 1$ and it is taken into account that $\beta_0<0$. 
Also, simple estimations show that typical $q\leq \mu\exp(-n)$. 

 Asymptotic series~\re{deltamIR} sometimes can
 be summed using the Borel transform. Formally, the Borel transform
of a series $f(\alpha)=\alpha\sum_{n=0}^{\infty}c_{n}\alpha^{n},$
with respect to $\alpha$  is defined as 
$B[f](t)=\sum_{n=0}^{\infty}c_{n}t^{n}/n!$.
If this Borel series converges, then the integral $I[f]=\int_{0}^{\infty}e^{-t/\alpha}B[f](t)\,dt\,\,\,$
gives the Borel sum of the original series. 

 The corresponding Borel transform of Eq.~\re{deltamIR} in respect to $(-\beta_0 \alpha(\mu)) $ is 
\begin{eqnarray}
B[\Delta m_{IR}](t)&=&-\frac{\mu\,4\pi C_{F}}{\beta_0}\sum_{n=0}^\infty (-1)^n\frac{t^n}{n!}\int_0^1 dx\, \left(\frac{x^2}{x^2+a_g^2}\right)^{n+1}
\ln^{n} x^2
\nonumber\\
&=&-\frac{\mu\,4\pi C_{F}}{\beta_0}\int_0^1 dx\,\left(\frac{x^2}{x^2+a_g^2}\right)\sum_{n=0}^\infty \frac{t^n}{n!}\left(-\frac{x^2}{x^2+a_g^2}\log x^2\right)^n
\nonumber\\
&=&-\frac{\mu\,4\pi C_{F}}{\beta_0}\int_0^1 dx\,\left(\frac{x^2}{x^2+a_g^2}\right)\exp\left(\frac{-t\,x^2}{x^2+a_g^2}\ln x^2\right).
\label{borel}\end{eqnarray}
The Eq.~(\ref{borel}) define the function $B[\Delta m](t)$ without singularities at least at any positive $t$.
So, we may conclude
that in ILM there is no IR renormalons in $\Delta m$.

For massless gluons $a_g=0$ the Borel transform of $\Delta m_{IR}$ is 
\begin{eqnarray}
B[\Delta m_{IR,a_g=0}](t)=-2\frac{\mu\,4\pi C_{F}}{\beta_0} \int_0^1 dx\,x^{-2t}
=-2\frac{\mu\,4\pi C_{F}}{\beta_0}\frac{1}{1-2t}
\label{borel1}\end{eqnarray}
and have the pole $t=1/2$. This pole correspond IR renormalon, which inhibits evaluation of
 $\Delta m_{IR,a_g=0}$ using inverse Borel transform (see recent paper~\cite{Beneke:2021lkq} and references therein). 
  
In ILM situation is much more comfortable, we may restore $\Delta m_{IR}$
by the calculation of the Borel integral
\begin{eqnarray}
I[\Delta m_{IR}] = -\frac{\mu\,4\pi C_{F}}{\beta_0} \int_0^\infty dt\, \int_0^1 dx\, \frac{x^2}{x^2+a_g^2}\exp\left(\frac{-t\,x^2}{x^2+a_g^2}\ln x^2
+\frac{t}{\beta_0\alpha(\mu)}\right).
\end{eqnarray}
Since the integrand of $I[\Delta m_{IR}]$ has no poles on $0<t<\infty$, we may change the order of the integration and make the integration on $t$ first, which gives
\begin{equation}
I[\Delta m_{IR}] = -\mu\,\alpha(\mu)\,4\pi\, C_{F}\int_0^1 dx\, \frac{x^2}{x^2+a_g^2-\beta_0\alpha(\mu)x^2\ln x^2}, 
\end{equation}
and further integration can be done numerically for any given values of $\alpha_s(\mu)$.

The same conclusion can be made about one-gluon exchange potential
$V(r)$ for colorless $Q\bar{Q}$ with account of radiative corrections
\begin{eqnarray}
V(r)=-4\pi\alpha_{s}(\mu)C_{F}\int\frac{d^{3}q}{(2\pi)^{3}}\exp({i\vec{q}\,\vec{r}}\,)\,
\bar{S}_{44}(q)\left(1-\frac{q^{2}}{q^{2}+M_{g}^{2}(q)}\pi(q)\right)^{-1},
\end{eqnarray}
since again at IR region the typical integrals will be the same as
shown at Eq.~\re{typical}.

\section{Conclusion}

We see from calculations above that in ILM framework it is safe to
use the pole heavy quark mass $m_{Q}$ and the perturbative potential
$V(r)$ for $Q\bar{Q}$-oniums, since there are no IR renormalons.
We plan to calculate perturbatively in ILM  the total energy of $Q\bar{Q}$
color singlet system $E(r)=2m_{Q}+V(r)$, since there is essential
cancellation of  IR region contributions to the $2m_{Q}$ and $V(r)$,
which is improving the convergence of perturbation  series in $\alpha_{s}$.

\section*{Acknowledgements}
M.M. is thankful to Marat Siddikov for the useful
and helpful communications.

\bibliography{refs2}

\begin{thebibliography}{10}
\expandafter\ifx\csname url\endcsname\relax
  \def\url#1{\texttt{#1}}\fi
\expandafter\ifx\csname urlprefix\endcsname\relax\def\urlprefix{URL }\fi
\expandafter\ifx\csname href\endcsname\relax
  \def\href#1#2{#2} \def\path#1{#1}\fi

\bibitem{Belavin:1975fg}
A.~A. Belavin, A.~M. Polyakov, A.~S. Schwartz, Y.~S. Tyupkin, {Pseudoparticle
  Solutions of the Yang-Mills Equations}, Phys. Lett. B 59 (1975) 85--87.
\newblock \href {https://doi.org/10.1016/0370-2693(75)90163-X}
  {\path{doi:10.1016/0370-2693(75)90163-X}}.

\bibitem{Pol}
A.~M. Polyakov, {Quark Confinement and Topology of Gauge Groups}, Nucl. Phys. B
  120 (1977) 429--458.
\newblock \href {https://doi.org/10.1016/0550-3213(77)90086-4}
  {\path{doi:10.1016/0550-3213(77)90086-4}}.

\bibitem{tH}
G.~'t~Hooft, {Symmetry Breaking Through Bell-Jackiw Anomalies}, Phys. Rev.
  Lett. 37 (1976) 8--11.
\newblock \href {https://doi.org/10.1103/PhysRevLett.37.8}
  {\path{doi:10.1103/PhysRevLett.37.8}}.

\bibitem{tH1}
G.~'t~Hooft, {Computation of the Quantum Effects Due to a Four-Dimensional
  Pseudoparticle}, Phys. Rev. D 14 (1976) 3432--3450, [Erratum: Phys.Rev.D 18,
  2199 (1978)].
\newblock \href {https://doi.org/10.1103/PhysRevD.14.3432}
  {\path{doi:10.1103/PhysRevD.14.3432}}.

\bibitem{FJR1976}
L.~D. Faddeev, {In Search for Multidimensional Solitons}, in: {4th
  International Conference on Nonlocal Quantum Field Theory, JINR Dubna}, 1976,
  pp. 207--223.

\bibitem{Jackiw:1976pf}
R.~Jackiw, C.~Rebbi, {Vacuum Periodicity in a Yang-Mills Quantum Theory}, Phys.
  Rev. Lett. 37 (1976) 172--175.
\newblock \href {https://doi.org/10.1103/PhysRevLett.37.172}
  {\path{doi:10.1103/PhysRevLett.37.172}}.

\bibitem{Schafer:1996wv}
T.~Sch\"afer, E.~V. Shuryak, {Instantons in QCD}, Rev. Mod. Phys. 70 (1998)
  323--426.
\newblock \href {http://arxiv.org/abs/hep-ph/9610451}
  {\path{arXiv:hep-ph/9610451}}, \href
  {https://doi.org/10.1103/RevModPhys.70.323}
  {\path{doi:10.1103/RevModPhys.70.323}}.

\bibitem{shuryak2018}
E.~Shuryak, {Lectures on nonperturbative QCD ( Nonperturbative Topological
  Phenomena in QCD and Related Theories)} (12 2018).
\newblock \href {http://arxiv.org/abs/1812.01509} {\path{arXiv:1812.01509}}.

\bibitem{Diakonov:2002fq}
D.~Diakonov, {Instantons at work}, Prog. Part. Nucl. Phys. 51 (2003) 173--222.
\newblock \href {http://arxiv.org/abs/hep-ph/0212026}
  {\path{arXiv:hep-ph/0212026}}, \href
  {https://doi.org/10.1016/S0146-6410(03)90014-7}
  {\path{doi:10.1016/S0146-6410(03)90014-7}}.

\bibitem{Chu:1994vi}
M.~C. Chu, J.~M. Grandy, S.~Huang, J.~W. Negele, {Evidence for the role of
  instantons in hadron structure from lattice QCD}, Phys. Rev. D 49 (1994)
  6039--6050.
\newblock \href {http://arxiv.org/abs/hep-lat/9312071}
  {\path{arXiv:hep-lat/9312071}}, \href
  {https://doi.org/10.1103/PhysRevD.49.6039}
  {\path{doi:10.1103/PhysRevD.49.6039}}.

\bibitem{Negele:1998ev}
J.~W. Negele, {Instantons, the QCD vacuum, and hadronic physics}, Nucl. Phys. B
  Proc. Suppl. 73 (1999) 92--104.
\newblock \href {http://arxiv.org/abs/hep-lat/9810053}
  {\path{arXiv:hep-lat/9810053}}, \href
  {https://doi.org/10.1016/S0920-5632(99)85010-5}
  {\path{doi:10.1016/S0920-5632(99)85010-5}}.

\bibitem{DeGrand:2001tm}
T.~A. DeGrand, {Short distance current correlators: Comparing lattice
  simulations to the instanton liquid}, Phys. Rev. D 64 (2001) 094508.
\newblock \href {http://arxiv.org/abs/hep-lat/0106001}
  {\path{arXiv:hep-lat/0106001}}, \href
  {https://doi.org/10.1103/PhysRevD.64.094508}
  {\path{doi:10.1103/PhysRevD.64.094508}}.

\bibitem{Faccioli:2003qz}
P.~Faccioli, T.~A. DeGrand, {Evidence for instanton induced dynamics, from
  lattice QCD}, Phys. Rev. Lett. 91 (2003) 182001.
\newblock \href {http://arxiv.org/abs/hep-ph/0304219}
  {\path{arXiv:hep-ph/0304219}}, \href
  {https://doi.org/10.1103/PhysRevLett.91.182001}
  {\path{doi:10.1103/PhysRevLett.91.182001}}.

\bibitem{Millo:2011zn}
R.~Millo, P.~Faccioli, {Computing the Effective Hamiltonian of Low-Energy
  Vacuum Gauge Fields}, Phys. Rev. D 84 (2011) 034504.
\newblock \href {http://arxiv.org/abs/1105.2163} {\path{arXiv:1105.2163}},
  \href {https://doi.org/10.1103/PhysRevD.84.034504}
  {\path{doi:10.1103/PhysRevD.84.034504}}.

\bibitem{Diakonov:1987ty}
D.~Diakonov, V.~Y. Petrov, P.~V. Pobylitsa, {A Chiral Theory of Nucleons},
  Nucl. Phys. B 306 (1988) 809.
\newblock \href {https://doi.org/10.1016/0550-3213(88)90443-9}
  {\path{doi:10.1016/0550-3213(88)90443-9}}.

\bibitem{Diakonov:1985eg}
D.~Diakonov, V.~Y. Petrov, {A Theory of Light Quarks in the Instanton Vacuum},
  Nucl. Phys. B 272 (1986) 457--489.
\newblock \href {https://doi.org/10.1016/0550-3213(86)90011-8}
  {\path{doi:10.1016/0550-3213(86)90011-8}}.

\bibitem{Goeke:2007bj}
K.~Goeke, M.~M. Musakhanov, M.~Siddikov, {Low energy constants of chi PT from
  the instanton vacuum model}, Phys. Rev. D 76 (2007) 076007.
\newblock \href {http://arxiv.org/abs/0707.1997} {\path{arXiv:0707.1997}},
  \href {https://doi.org/10.1103/PhysRevD.76.076007}
  {\path{doi:10.1103/PhysRevD.76.076007}}.

\bibitem{Kim:2004hd}
H.-C. Kim, M.~Musakhanov, M.~Siddikov, {Magnetic susceptibility of the QCD
  vacuum}, Phys. Lett. B 608 (2005) 95--106.
\newblock \href {http://arxiv.org/abs/hep-ph/0411181}
  {\path{arXiv:hep-ph/0411181}}, \href
  {https://doi.org/10.1016/j.physletb.2004.12.080}
  {\path{doi:10.1016/j.physletb.2004.12.080}}.

\bibitem{Kim:2005jc}
H.-C. Kim, M.~M. Musakhanov, M.~Siddikov, {Meson-loop contributions to the
  quark condensate from the instanton vacuum}, Phys. Lett. B 633 (2006)
  701--709.
\newblock \href {http://arxiv.org/abs/hep-ph/0508211}
  {\path{arXiv:hep-ph/0508211}}, \href
  {https://doi.org/10.1016/j.physletb.2005.11.054}
  {\path{doi:10.1016/j.physletb.2005.11.054}}.

\bibitem{Kraan:1998kp}
T.~C. Kraan, P.~van Baal, {Exact T duality between calorons and Taub - NUT
  spaces}, Phys. Lett. B 428 (1998) 268--276.
\newblock \href {http://arxiv.org/abs/hep-th/9802049}
  {\path{arXiv:hep-th/9802049}}, \href
  {https://doi.org/10.1016/S0370-2693(98)00411-0}
  {\path{doi:10.1016/S0370-2693(98)00411-0}}.

\bibitem{Kraan:1998pm}
T.~C. Kraan, P.~van Baal, {Periodic instantons with nontrivial holonomy}, Nucl.
  Phys. B 533 (1998) 627--659.
\newblock \href {http://arxiv.org/abs/hep-th/9805168}
  {\path{arXiv:hep-th/9805168}}, \href
  {https://doi.org/10.1016/S0550-3213(98)00590-2}
  {\path{doi:10.1016/S0550-3213(98)00590-2}}.

\bibitem{Lee:1998bb}
K.-M. Lee, C.-h. Lu, {SU(2) calorons and magnetic monopoles}, Phys. Rev. D 58
  (1998) 025011.
\newblock \href {http://arxiv.org/abs/hep-th/9802108}
  {\path{arXiv:hep-th/9802108}}, \href
  {https://doi.org/10.1103/PhysRevD.58.025011}
  {\path{doi:10.1103/PhysRevD.58.025011}}.

\bibitem{Diakonov:2009jq}
D.~Diakonov, {Topology and confinement}, Nucl. Phys. B Proc. Suppl. 195 (2009)
  5--45.
\newblock \href {http://arxiv.org/abs/0906.2456} {\path{arXiv:0906.2456}},
  \href {https://doi.org/10.1016/j.nuclphysbps.2009.10.010}
  {\path{doi:10.1016/j.nuclphysbps.2009.10.010}}.

\bibitem{Liu:2015ufa}
Y.~Liu, E.~Shuryak, I.~Zahed, {Confining dyon-antidyon Coulomb liquid model.
  I.}, Phys. Rev. D 92~(8) (2015) 085006.
\newblock \href {http://arxiv.org/abs/1503.03058} {\path{arXiv:1503.03058}},
  \href {https://doi.org/10.1103/PhysRevD.92.085006}
  {\path{doi:10.1103/PhysRevD.92.085006}}.

\bibitem{Liu:2015jsa}
Y.~Liu, E.~Shuryak, I.~Zahed, {Light quarks in the screened dyon-antidyon
  Coulomb liquid model. II.}, Phys. Rev. D 92~(8) (2015) 085007.
\newblock \href {http://arxiv.org/abs/1503.09148} {\path{arXiv:1503.09148}},
  \href {https://doi.org/10.1103/PhysRevD.92.085007}
  {\path{doi:10.1103/PhysRevD.92.085007}}.

\bibitem{Musakhanov:1998wp}
M.~Musakhanov, {Improved effective action for light quarks beyond chiral
  limit}, Eur. Phys. J. C 9 (1999) 235--243.
\newblock \href {http://arxiv.org/abs/hep-ph/9810295}
  {\path{arXiv:hep-ph/9810295}}, \href {https://doi.org/10.1007/s100529900017}
  {\path{doi:10.1007/s100529900017}}.

\bibitem{Musakhanov:2002vu}
M.~Musakhanov, {Current mass dependence of the quark condensate in instanton
  vacuum}, Nucl. Phys. A 699 (2002) 340--343.
\newblock \href {https://doi.org/10.1016/S0375-9474(01)01516-0}
  {\path{doi:10.1016/S0375-9474(01)01516-0}}.

\bibitem{Musakhanov:2002xa}
M.~M. Musakhanov, H.-C. Kim, {A Test of the instanton vacuum with low-energy
  theorems of the axial anomaly}, Phys. Lett. B 572 (2003) 181--188.
\newblock \href {http://arxiv.org/abs/hep-ph/0206233}
  {\path{arXiv:hep-ph/0206233}}, \href
  {https://doi.org/10.1016/j.physletb.2003.08.022}
  {\path{doi:10.1016/j.physletb.2003.08.022}}.

\bibitem{Beneke:2021lkq}
M.~Beneke, {Pole mass renormalon and its ramifications, } (8 2021).
\newblock \href {http://arxiv.org/abs/2108.04861} {\path{arXiv:2108.04861}},
  \href {https://doi.org/10.1140/epjs/s11734-021-00268-w}
  {\path{doi:10.1140/epjs/s11734-021-00268-w}}.

\bibitem{hutter}
M.~Hutter, {Gluon mass from instantons, } (11 1993).
\newblock \href {http://arxiv.org/abs/hep-ph/9501335}
  {\path{arXiv:hep-ph/9501335}}.

\bibitem{Musakhanov:2017erp}
M.~Musakhanov, O.~Egamberdiev, {Dynamical gluon mass in the instanton vacuum
  model}, Phys. Lett. B 779 (2018) 206--209.
\newblock \href {http://arxiv.org/abs/1706.06270} {\path{arXiv:1706.06270}},
  \href {https://doi.org/10.1016/j.physletb.2018.01.080}
  {\path{doi:10.1016/j.physletb.2018.01.080}}.

\bibitem{Musakhanov:2020hvk}
M.~Musakhanov, N.~Rakhimov, U.~T. Yakhshiev, {Heavy quark correlators in an
  instanton liquid model with perturbative corrections}, Phys. Rev. D 102~(7)
  (2020) 076022.
\newblock \href {http://arxiv.org/abs/2006.01545} {\path{arXiv:2006.01545}},
  \href {https://doi.org/10.1103/PhysRevD.102.076022}
  {\path{doi:10.1103/PhysRevD.102.076022}}.

\bibitem{Broadhurst:1994se}
D.~J. Broadhurst, A.~G. Grozin, {Matching QCD and HQET heavy - light currents
  at two loops and beyond}, Phys. Rev. D 52 (1995) 4082--4098.
\newblock \href {http://arxiv.org/abs/hep-ph/9410240}
  {\path{arXiv:hep-ph/9410240}}, \href
  {https://doi.org/10.1103/PhysRevD.52.4082}
  {\path{doi:10.1103/PhysRevD.52.4082}}.

\bibitem{Beneke:1994qe}
M.~Beneke, V.~M. Braun, {Naive nonAbelianization and resummation of fermion
  bubble chains}, Phys. Lett. B 348 (1995) 513--520.
\newblock \href {http://arxiv.org/abs/hep-ph/9411229}
  {\path{arXiv:hep-ph/9411229}}, \href
  {https://doi.org/10.1016/0370-2693(95)00184-M}
  {\path{doi:10.1016/0370-2693(95)00184-M}}.

\end{thebibliography}

\end{document}